\documentclass{aa}
\usepackage{graphicx,color,subfigure,multirow}
\usepackage{txfonts}
\usepackage{verbatim}
\graphicspath{{./figures/}}
\usepackage{natbib,twoopt}
\usepackage[breaklinks=true]{hyperref}
\hypersetup{
	colorlinks=true, 
	urlcolor=blue,
	linkcolor=blue
}
\bibpunct{(}{)}{;}{a}{}{,}

\makeatletter
\newcommand{\bibnote}[2]{\global\@namedef{#1note}{#2}}
\newcommand{\biblink}[2]{\global\@namedef{#1link}{#2}}
\makeatother
\makeatletter
\protected\def\stonyslink{%
	\def\hyper@linkstart##1##2{}\let\hyper@linkend\@empty}
\newcommandtwoopt{\citeads}[3][][]{%
	\href{http://adsabs.harvard.edu/abs/#3}%
	{\stonyslink \citealp[#1][#2]{#3}}%
	\biblink{#3}{\href{http://adsabs.harvard.edu/abs/#3}{ADS}}}
\newcommandtwoopt{\citepads}[3][][]{%
	\href{http://adsabs.harvard.edu/abs/#3}%
	{\stonyslink \citep[#1][#2]{#3}}%
	\biblink{#3}{\href{http://adsabs.harvard.edu/abs/#3}{ADS}}}
\newcommandtwoopt{\citetads}[3][][]{%
	\href{http://adsabs.harvard.edu/abs/#3}%
	{\stonyslink \citet[#1][#2]{#3}}%
	\biblink{#3}{\href{http://adsabs.harvard.edu/abs/#3}{ADS}}}
\newcommandtwoopt{\citeyearads}[3][][]{%
	\href{http://adsabs.harvard.edu/abs/#3}%
	{\stonyslink \citeyear[#1][#2]{#3}}%
	\biblink{#3}{\href{http://adsabs.harvard.edu/abs/#3}{ADS}}}
\makeatother

\begin{document}

\title{Acceleration of planetary migration: Resonance crossing and planetesimal ring}
\author{
    Hailiang Li\inst{1}
    \and
	Li-Yong Zhou\inst{2}\fnmsep\inst{3}
	\and
    Xiaoping Zhang\inst{1}
	}
\authorrunning{Li et al.}
\institute{State Key Laboratory of Lunar and Planetary Sciences, Macau University of Science and Technology, Macau 999078, China\\
            \href{mailto:xpzhang@must.edu.mo}{xpzhang@must.edu.mo}
            \and
            School of Astronomy and Space Science, Nanjing University, 163 Xianlin Avenue, Nanjing 210046, China
	      \and
	    Key Laboratory of Modern Astronomy and Astrophysics in Ministry of Education, Nanjing University, China
	}
\date{}

\abstract{Planetary migration is a crucial stage in the early Solar System, explaining many observational phenomena and providing constraints on details related to the Solar System's origins. This paper aims to investigate the acceleration during planetary migration in detail using numerical simulations, delving deeper into the early Solar System's preserved information. We confirm that planetary migration is a positive feedback process: the faster the migration, the more efficient the consumption of planetesimals; once the migration slows down, Neptune clears the surrounding space, making further migration more difficult to sustain. Quantitatively, a tenfold increase in the migration rate corresponds to a reduction of approximately 30\% in the mass of planetesimals consumed to increase per unit of angular momentum of Neptune. We also find that Neptune's final position is correlated with the initial surface density of planetesimals at that location, suggesting that the disk density at 30\,au was approximately 0.009\,$M_{\oplus}/au^2$ in the early Solar System. Furthermore, we identify two mechanisms that can accelerate planetary migration. The first is mean motion resonance between Uranus and Neptune: migration acceleration will be triggered whenever these two giant planets cross their major mean motion resonance. The second mechanism is the ring structure within the planetesimal disk, as the higher planetesimal density in this region can provide the material support necessary for migration acceleration. Our research indicates that Neptune in the current Solar System occupies a relatively delicate position. In case Neptune crossed the 1:2 resonance with Uranus, it could have migrated to a much more distant location. Our results demonstrate that giant planet instability is fundamentally required; otherwise, reconstructing the migration histories of Uranus and Neptune would yield physically implausible orbital configurations. Furthermore, even without introducing the giant planet instability, under the influence of the positive feedback mechanism, the evolution of the Solar System to its current configuration might still be a stochastic outcome rather than an inevitable consequence.
}

\maketitle{}

\section{Introduction}
Planetary migration is a phenomenon in the early formation and evolution of the Solar System, which was first proposed by \citetads{Fernandez1984}. In the classical model of planetary migration, the exchange of angular momentum between planets and small bodies is the primary driver of migration. As the small bodies gradually deplete, the rate of planetary migration slows down. It is generally believed that the migration rate follows an approximately exponential decay pattern \citepads{Malhotra1995}.

The process of planetary migration has profoundly shaped the dynamical structure of the current Solar System. People have known that planetary migration could explain Pluto's high orbital eccentricity for decades \citepads{Malhotra1993,Malhotra1995}. With the discovery of numerous trans-Neptunian objects (TNOs), the planetary migration model was further found to account for the origins of various TNO populations \citepads[e.g.,][]{Levison2008,Morbidelli2020}. To date, the most successful framework for understanding planetary migration is the Nice model, in which the planetary migration has been incorporated into sophisticated scenarios that provide successful explanations for various features in the Solar System \citepads[see][for a review]{Nesvorny2018}. A critical stage in the Nice model is planet instability, in which the planets cross the major mean motion resonance (MMR) causing violent changes in planets' orbits \citepads[e.g.,][]{Tsiganis2005} or even loss(es) of planet(s) from the young Solar System \citepads[e.g.,][]{Nesvorny2012}. Building on the framework, subsequent investigations have systematically examined various initial giant planet configurations, analyzing different resonant chains, instability trigger timing, and the number of planets \citepads[e.g.,][]{Nesvorny2012,Pierens2014,Deienno2017,Ribeiro2020,Clement2021}.

Previous simulations on planetary migration have shown that, in a single-planet model, the interaction between the planet and planetesimals typically causes the planet to migrate inward \citepads{Kirsh2009}. Because nearly all angular momentum loss results from planetesimals being scattered out of the Solar System. While in the context of the early Solar System with multiple giant planets present, the overall effect of planetary migration is to make the planets scatter apart and move away from each other \citepads[e.g.,][]{Hahn1999,Gomes2004}.

However, certain mechanisms can also enable outward migration in a single-planet scenario. When the density of the protoplanetary disk is sufficiently high, planetary migration can lead to an asymmetric distribution of planetesimals on either side of the planet, extending the migration to a greater distance. This self-sustaining migration is referred to as ``runaway migration'' \citepads{Ida2000}. On the other hand, in the earlier stages of the Solar System when the gas in the protoplanetary disk had not yet dissipated, as gas travels along a horseshoe trajectory past a migrating planet, the torque exerted by the gas on the planet encourages further migration. This positive feedback mechanism is known as Type \uppercase\expandafter{\romannumeral3} migration \citepads{Masset2003, Peplinski2008}.

It has often been assumed that the density distribution of the planetesimals is smooth. However, recent studies increasingly suggest that the planetesimal disk likely contains ring structures. The locations of these rings are related to the temperature distribution of the early Solar System, corresponding to the snow lines or sublimation lines of specific materials, such as water and silicates \citepads[e.g.,][]{Izidoro2022}. Using high-resolution imaging techniques, the Atacama Large Millimeter/submillimeter Array (ALMA) has identified distinct ring-like structures within the protoplanetary disk of HL Tauri \citepads{ALMA2015}. These ring structures are thought to be closely linked to the planet formation process in protoplanetary disks \citepads{Lichtenberg2021, Izidoro2022, Morbidelli2022, Batygin2023}. Naturally, it can be inferred that these ring structures could have persisted from the protoplanetary disk into the later planetesimal disk, thereby playing a significant role in the process of planetary migration.

Building on the conclusions of the aforementioned studies, it is reasonable to speculate that Neptune might have encountered a ring structure where the material density is higher during its migration. This encounter may have triggered the runaway migration for Neptune. Through a positive feedback mechanism, Neptune's migration rate increased. Subsequently, Neptune passed through this planetesimal ring and eventually settled at its current position. To validate this conjecture, we aim to recreate this physical process using numerical simulations in this paper. 

The primary objective of this study is not to fully explain the current Solar System architecture or TNO configurations, but rather to test the dynamical effects of Neptune's migration models under various conditions. Therefore, we have not specifically incorporated a giant planet instability process in this investigation. This approach may be oversimplified; nevertheless, it enables us to derive insights into broader applicability. For example, this study may be interpreted as representing a post-instability migration phase to some extent.

We introduce the research methods in Section 2. In Section 3, we explore the relationship between planetary migration and planetesimal density within a smooth planetesimal disk. Section 4 focuses on testing the acceleration of Neptune as it passes through a planetesimal ring. The final section comprises the conclusion and a discussion.

\section{Methods}

In designing the planetary migration model, we employed an outer Solar System model with four gas giants. It is generally agreed that the initial positions of the planets were more compact than their current ones, and research has demonstrated that planets would evolve into resonant chains under the influence of gas disk \citepads[][]{Morbidelli2007a,Morbidelli2007b}. However, since we are not to delve into the influence of the giant planet instability, we decided to refer to earlier studies \citepads[e.g.,][]{Malhotra1995, Hahn1999, Gomes2004}. In this paper, we denote the initial positions of the four giant planets as $A_{ini}$ = [$a_J$, $a_S$, $a_U$, $a_N$], where the subscripts J, S, U, N represent Jupiter, Saturn, Uranus, and Neptune, respectively. In the simulations of Section 3.1, we adopted the initial positions used by \citetads{Gomes2004}; namely, $[5.45\,au, 8.7\,au, 15.5\,au, 17.8\,au]$. In Section 3.4, based on the relative positions of Uranus and Neptune, we explored another set of initial conditions as $A_{ini} = [5.45\,au, 8.8\,au, 19.8\,au, 23\,au]$, which was used in the calculations in Section 4.2. Additionally, the initial masses of the giant planets were set to match those of the present-day Solar System, with their inclinations and eccentricities both assumed to be zero at the beginning of the simulations.

The total mass of the planetesimal disk ($M_{disk}$) is another critical parameter. Previous studies generally assumed the $M_{disk}$ of several tens of Earth masses ($M_{\oplus}$), as larger values would produce excessive damping of planetary eccentricities \citepads[e.g.,][]{Tsiganis2005,Nesvorny2012}, while \citetads{Nesvorny2012} suggested a planetesimal disk with $M_{disk} \approx 20\,M_{\oplus}$, and truncated at 30\,au. In this article, we adopt a more extended disk (with the outer boundary, $a_{out}$, typically set to 50\,au) to thoroughly characterize Neptune's migration behavior. In Section 3 we selected several different initial $M_{disk}$ from 30\,$M_{\oplus}$ to 60\,$M_{\oplus}$ for testing. It should be noted that although the $M_{disk}$ is larger than the value suggested by \citetads{Nesvorny2012}, the surface density is comparable to that of a disk with $M_{disk} \approx 20\,M_{\oplus}$ but $a_{out}=30\,au$. In Section 3.1, the total number of small bodies is $10^4$, and all small bodies have identical masses (i.e., ranging from 0.003\,$M_{\oplus}$ to 0.006\,$M_{\oplus}$) within a single simulation.

Traditionally, researchers have assumed that the planetesimal disk has a smooth mass distribution, whereby the surface density, $\Sigma$, of the planetesimal disk follows a power-law distribution, $\Sigma \propto r^{-\alpha}$. Here, $r$ represents the heliocentric distance and $\alpha$ is the parameter describing the distribution. In this paper, we refer to this type of planetesimal disk with a smooth mass density distribution as a plain planetesimal disk. Typically, the value of $\alpha$ is chosen to be 1 or 1.5 \citepads[e.g.,][]{Hahn1999, Gomes2003, Gomes2004, Kirsh2009, Nesvorny2012, Nesvorny2015a, Deienno2017, Clement2021}. \citetads{Gomes2004} noted that a sharper density distribution (e.g., $\Sigma \propto r^{-4}$) may help explain why Neptune could stop at a distance of 30\,au. Therefore, in the tests of Section 3, we selected five different density distributions with $\alpha=1, 1.5, 2, 3, 4$, covering typical values of $\alpha$.

A key question concerns the distance between the inner boundary of the planetesimal disk and the outermost ice giant, which is closely tied to the timing of planetary migration and the giant planet instability \citepads[][]{Nesvorny2012}. Numerous previous studies have conducted meaningful investigations into this issue, testing separations between the planetesimal disk and Neptune ranges from several tenths of an astronomical unit to a few astronomical units\citepads[e.g.,][]{Tsiganis2005, Levison2011, Nesvorny2012, Deienno2017, Ribeiro2020}. Recent work suggests an early instability\citepads[e.g.,][]{Clement2018, Nesvorny2018b, Clement2019, Ribeiro2020}, implying that the separation should have been relatively small. In this study, we adopted a narrower gap configuration, setting the disk's inner edge, $a_{in}$, as $a_N+0.2\,au$. Although we did not account for giant planet instability, such an idealized assumption is acceptable because this setup will trigger planetary migration at the onset of the simulation. With these specifications, the overall profile of the planetesimal disk can be determined, and the surface density ($\Sigma$) at any heliocentric distance (r) is given by

\begin{equation}
\Sigma(r)=
\left\{
\begin{aligned}
& \text{$ \frac{(2-\alpha)M_{disk}}{2\pi (a_{out}^{2-\alpha}-a_{in}^{2-\alpha})}r^{-\alpha} \qquad \text{for} \quad \alpha \neq 2 $}, \\
& \text{$ \frac{M_{disk}}{2\pi (\ln(a_{out})-\ln(a_{in}))}r^{-\alpha} \quad \text{for} \quad \alpha = 2 $ }. \\
\end{aligned}
\right.
\label{eq:surface_density}
\end{equation}

Based on Eq.~\ref{eq:surface_density}, the radial density is given by $2\pi r \Sigma(r)$. Assuming $a_{in}=18\,au$ and $a_{out}=50\,au$, the positions of the mass equipartition points for $\alpha=1, 1.5, 2, 3, 4$ were calculated to be $34.00\,au, 32.00\,au, 30.00\,au, 26.47\,au, 23.95\,au$, respectively.

It has been recognized that the eccentricities of main-belt asteroids follow a Rayleigh distribution \citepads[][]{Plummer1916,Beck1981,Malhotra2017}. Therefore, when designing a planetesimal disk, it is often assumed that both eccentricities and inclinations also follow a Rayleigh distribution \citepads[e.g.,][]{Ida1992,Kirsh2009,Capobianco2011,Clement2021}. In this paper, we adopt a dynamically cold planetesimal disk, whereby the initial eccentricity and inclination distributions of the planetesimals follow a Rayleigh distribution with parameters of $\sigma_e=2\,\sigma_i=0.002$, where $\sigma_e$ and $\sigma_i$ are the standard deviations of eccentricity and inclination, respectively.

To simulate planetary migration within a planetesimal disk, it is necessary to consider gravitational interactions between each planetesimal and the planets, which is computationally intensive. To achieve faster computation, we employed the GPU-based N-body integration code GENGA for our simulations \citepads{GENGA,GENGA2}. In our calculations, interactions between planetesimals are neglected. In Section 3.1, our simulations last a total of 500\,Myr to ensure that planetary migration have substantially completed.

In the simulations, planetesimals are removed for various reasons, such as collisions with the Sun or planets. In this study, we refer to the removal of planetesimals from the simulation as ``planetesimal consumption.'' The primary reason for planetesimal consumption is their being scattered out of the Solar System (heliocentric distance exceeding 1000\,au), with collisions accounting for only a minor fraction. Planetesimal consumption is closely related to close encounters experienced during Neptune's migration. Generally, as Neptune's orbit expands, planetesimals first approach Neptune and their orbits are then excited, turning them into Centaurs, and eventually being scattered out of the Solar System by Jupiter. This study will also explore in depth the relationship between planetary migration and planetesimal consumption.

\section{Migration in a plain planetesimal disk}

In this section, we first consider the simplest scenario in which Neptune migrates through a plain planetesimal disk. This approach aims to understand the fundamental patterns of Neptune's migration and investigate the factors influencing its migration behavior. Based on the initial conditions described in the previous section, we conducted a total of 25 simulations using $\alpha=1, 1.5, 2, 3, 4$ and $M_{disk}=30\,M_{\oplus}, 40\,M_{\oplus}, 45\,M_{\oplus}, 50\,M_{\oplus}, 60\,M_{\oplus}$.

\subsection{Neptune migration and MMR crossing}

\begin{figure}[!htb]
\centering
\resizebox{\hsize}{!}{\includegraphics{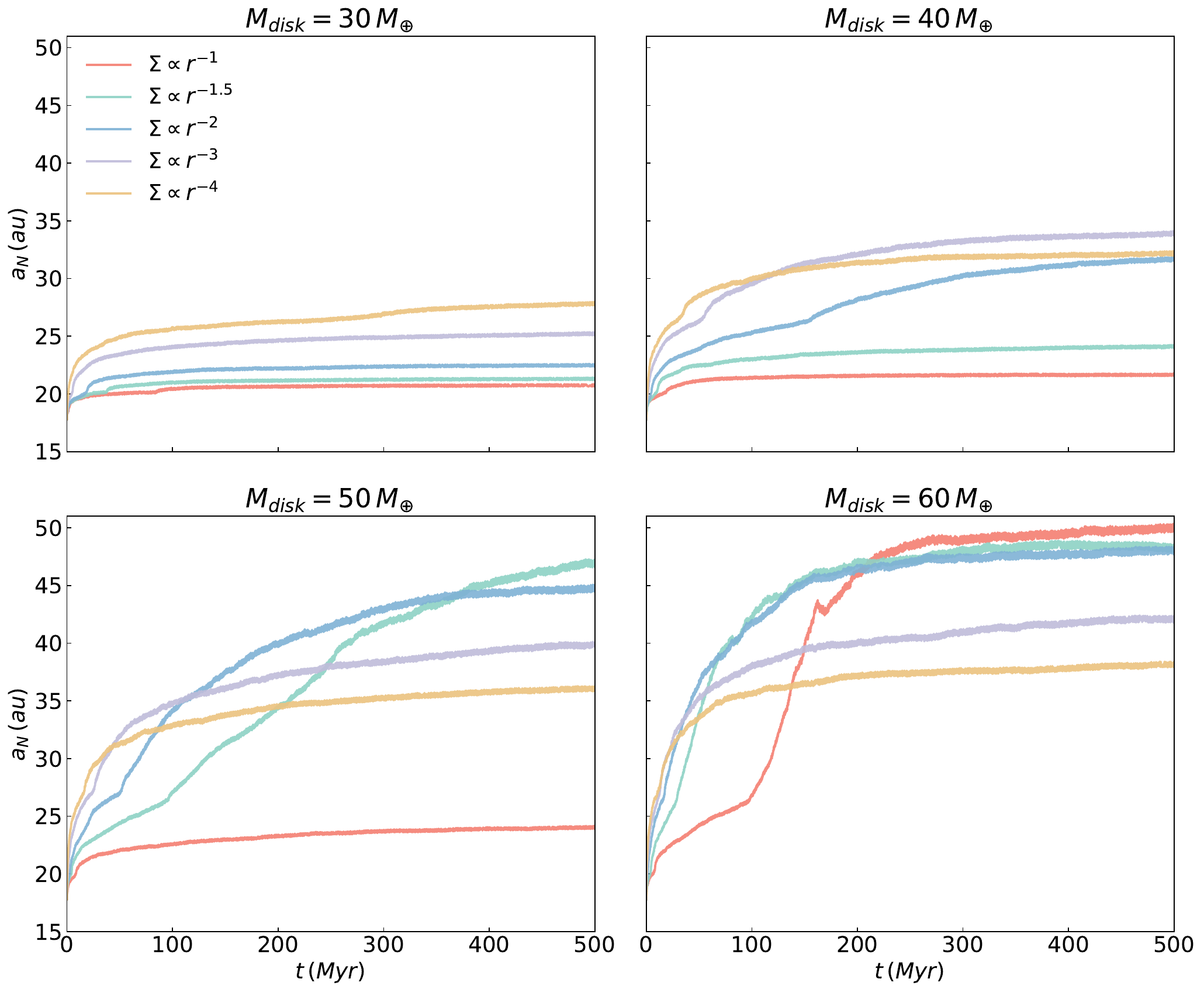}}
\caption{Evolution of $a_N$ in a plain planetesimal disk. The horizontal and vertical axes represent time and $a_N$, respectively. The four subplots correspond to different $M_{disk}$, with five different-colored lines representing various $\alpha$.}
\label{fig:Nep_mig}
\end{figure}

Notably, the current eccentricity of Jupiter serves as key evidence of giant planet instability and is used as a success criterion for migration simulations \citepads[e.g.,][]{Nesvorny2012, Deienno2017, Clement2021}. Since giant planet instability is not our primary focus, we have not attempted to reproduce the actual eccentricities of planets in this study. Under the interaction between the planets and the planetesimal disk, planetary eccentricities and mutual inclinations will be gradually damped, which is known as dynamical friction \citepads{Kokubo1995}. In these simulations, we monitored the orbital excitation of the giant planets and found that their orbits remained relatively stable, with eccentricities generally maintained below 0.01 and inclinations never exceeding $1^{\circ}$. Therefore, giant planet instability does not interfere with the migration process in this paper. In Fig.~\ref{fig:Nep_mig}, we present the evolution of $a_N$ to reflect the overall migration behavior.

Fig.~\ref{fig:Nep_mig} illustrates that different planetesimal distributions significantly influence the final position that Neptune can reach. Generally, as planetesimals are gradually depleted, Neptune's migration exhibits a damping trend. For a given $M_{disk}$, an appropriate density distribution can lead Neptune to achieve the farthest migration distance. Here we denote the $\alpha$ corresponding to the furthest migration as $\alpha_m$. For $M_{disk}=60\,M_{\oplus}$, the $\alpha_m$ will be 1. In this case, a shallower distribution of planetesimals facilitates Neptune's migration more effectively. However, this trend reverses for $M_{disk}=30\,M_{\oplus}$, where $\alpha_m=4$. For intermediate $M_{disk}$ of 40\,$M_{\oplus}$, 45\,$M_{\oplus}$, and 50\,$M_{\oplus}$, the observed $\alpha_m$ values are 3, 2, and 1.5, respectively.

Two mechanisms might suppress Neptune’s migration. When $\alpha>\alpha_m$, excessive mass concentrates in the inner planetesimal disk. Although this allows Neptune to migrate faster in the early stages, it also means that the outer region becomes too depleted, causing migration to stop prematurely. This aligns with the conclusion of \citetads{Gomes2004}: that a steeper density profile leads to earlier termination of Neptune’s migration.

The second mechanism occurs when $\alpha<\alpha_m$. In this case, Neptune’s migration is significantly slower from the outset compared to the migration under $\alpha_m$. Because a larger fraction of the mass is placed in the outer planetesimal disk, the less massive inner disk cannot sustain Neptune’s migration. Consequently, a substantial population of planetesimals is preserved in the outer disk, and never engages in the migration of Neptune.

We also note that in some previous studies \citepads[e.g.,][]{Nesvorny2012, Nesvorny2015a, Deienno2017}, Neptune could still migrate efficiently even when $\alpha=1$, eventually reaching the outer edge of the disk. This does not contradict our results, as the key difference lies in our choice of a more extended disk ($a_{out} = 50\,au$). This setup leads to a lower surface density in our simulations, and the inner disk lacks sufficient material to drive migration. On the other hand, most planetesimals are too distant to have strong gravitational interactions with Neptune. An additional factor may be that, in such an extended disk, Neptune captures more objects in its MMRs. The co-migration of these objects may act as a drag, impeding Neptune's migration.

\begin{figure}[!htb]
\centering
\resizebox{\hsize}{!}{\includegraphics{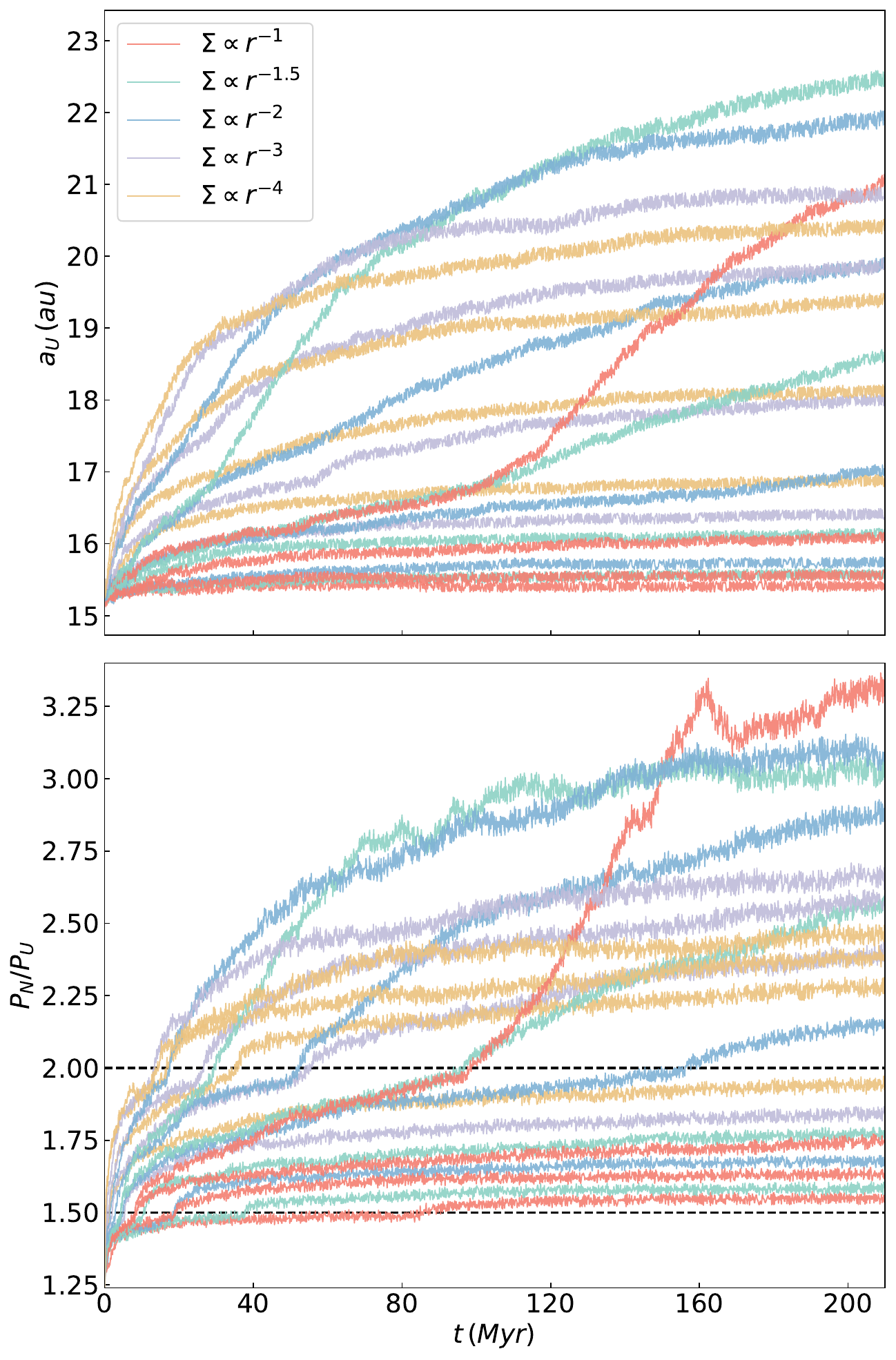}}
\caption{Acceleration of migration at crossing MMRs. The upper panel shows the evolution of $a_U$, while the lower panel displays the period ratio between Neptune and Uranus. Different $\alpha$ are represented by different colors. The $M_{disk}$ are not explicitly distinguished, but their relative positions are fixed, with four lines of the same color representing $M_{disk}$ of 30\,$M_{\oplus}$, 40\,$M_{\oplus}$, 50\,$M_{\oplus}$, and 60\,$M_{\oplus}$ from bottom to top. In the lower panel, two dashed black lines indicate positions where the period ratio between Neptune and Uranus is 2 and 1.5.}
\label{fig:PNU}
\end{figure}

In addition, in Fig.~\ref{fig:Nep_mig}, it is evident that an acceleration in migration often occurs when Neptune reaches approximately 26 to 27\,au. This acceleration is due to Uranus and Neptune crossing their 1:2 MMR. For convenience, we hereafter refer to the 1:2 MMR between Uranus and Neptune as $R_{2N:1U}$. In Fig.~\ref{fig:PNU}, we present the evolution of $a_U$ and the period ratio between Neptune and Uranus from 20 simulations to illustrate the impact of such resonance crossings on planetary migration. It is worth mentioning that a similar acceleration phenomenon has been observed by
\citetads{Gomes2004}, who suggested that this acceleration occurs when Neptune's 1:2 MMR extends beyond the outer boundary of the planetesimal disk (located at 50\,au). We attempted to adjust the $a_{out}$ to 45\,au or 55\,au, but the location of acceleration remained unchanged, which implies that the acceleration observed in Fig.~\ref{fig:Nep_mig} happens under a distinct mechanism.

In the lower panel of Fig.~\ref{fig:PNU}, it can be observed that as Neptune and Uranus migrate outward, their period ratio increases gradually. Each time the period ratio crosses 2, a small jump occurs, followed by a faster expansion compared to before. This corresponds to the migration acceleration of Neptune near 26$\sim$27\,au. We note that Neptune's location of the $R_{2N:1U}$ crossing is not fixed. In Fig.~\ref{fig:Nep_mig} where the initial position of planets are fixed, faster Neptune migration correlates with greater heliocentric distances at MMR crossing. This indicates that during faster migration, the ratio of the migration distances of Uranus and Neptune ($\Delta a_U/\Delta a_N$) will be slightly larger. More generally, Neptune's exact position when crossing the $R_{2N:1U}$ depends on the initial conditions, particularly the initial locations of Uranus and Neptune. In the real Solar System, Neptune has migrated beyond 30\,au without yet crossing the $R_{2N:1U}$; we investigate this further in the following text.

Additionally, when the period ratio between Neptune and Uranus crosses 1.5, another migration acceleration is observed, which corresponds to the crossing of the 2:3 MMR between Uranus and Neptune (hereafter referred to as $R_{3N:2U}$). Since $R_{3N:2U}$ occurs during the earlier, faster migration phase (with Neptune located near 20\,au), this acceleration is more pronounced in simulations with slower migration rates and shorter migration distances.

Similar effects can be observed in many previous works\citepads[e.g.,][]{Gomes2004, Tsiganis2005, Gomes2016, Nesvorny2018}, but past discussions on this acceleration have been little emphasized. Part of the reason is that the MMR between Neptune and Uranus is significantly weaker than the Jupiter-Saturn MMR. The MMR crossing between Jupiter and Saturn has been well studied as a potential triggering mechanism for giant planet instability \citepads[e.g.,][]{Tsiganis2005, Morbidelli2009}. 

In the case of $R_{2N:1U}$ crossing, we find that the resonance produces only transient eccentricity excitation ($e\sim0.02$) for Neptune and Uranus, with dynamical friction rapidly restoring their eccentricities to e < 0.01. Additionally, the orbits of planetesimals were modestly excited during resonance crossing, though the overall effect remained comparatively weak. Nevertheless, the crossing of $R_{2N:1U}$ can remarkably enhance Neptune's migration rate, accelerating it to several times faster than before, which may profoundly shape the orbital distribution of TNOs \citepads[e.g.,][]{Nesvorny2015b, Lawler2019, Morbidelli2020, Li2023}.

As we know, when the two giant planets cross the $R_{2N:1U}$, Neptune gains an outward migration acceleration, while $a_U$ (around 17\,au at that time) should theoretically experience a slight decrease to conserve angular momentum. However, as is shown in the upper panel of Fig.~\ref{fig:PNU}, the change in $a_U$ resulting from the resonance crossing is negligible ($< 0.1\,au$), indicating that Uranus did not lose much angular momentum. Instead, it continues to migrate outward alongside Neptune's acceleration in the subsequent evolution.

These phenomena suggest that the resonance crossing is only a relatively minor event in the planetary migration, and the subsequent rapid outward migration is still primarily driven by planetesimals. However, planetary migration under the influence of just the planetesimal disk tends to decay over time. The resonance crossing acts as a kind of ``trigger,'' giving Neptune a slight outward push that exposes it to a greater reservoir of planetesimals. Through the positive feedback mechanism, even this minor difference can dramatically amplify the total migration distance.

\subsection{Effect of planetesimal density on migration}

\begin{figure}[!htb]
\centering
\resizebox{\hsize}{!}{\includegraphics{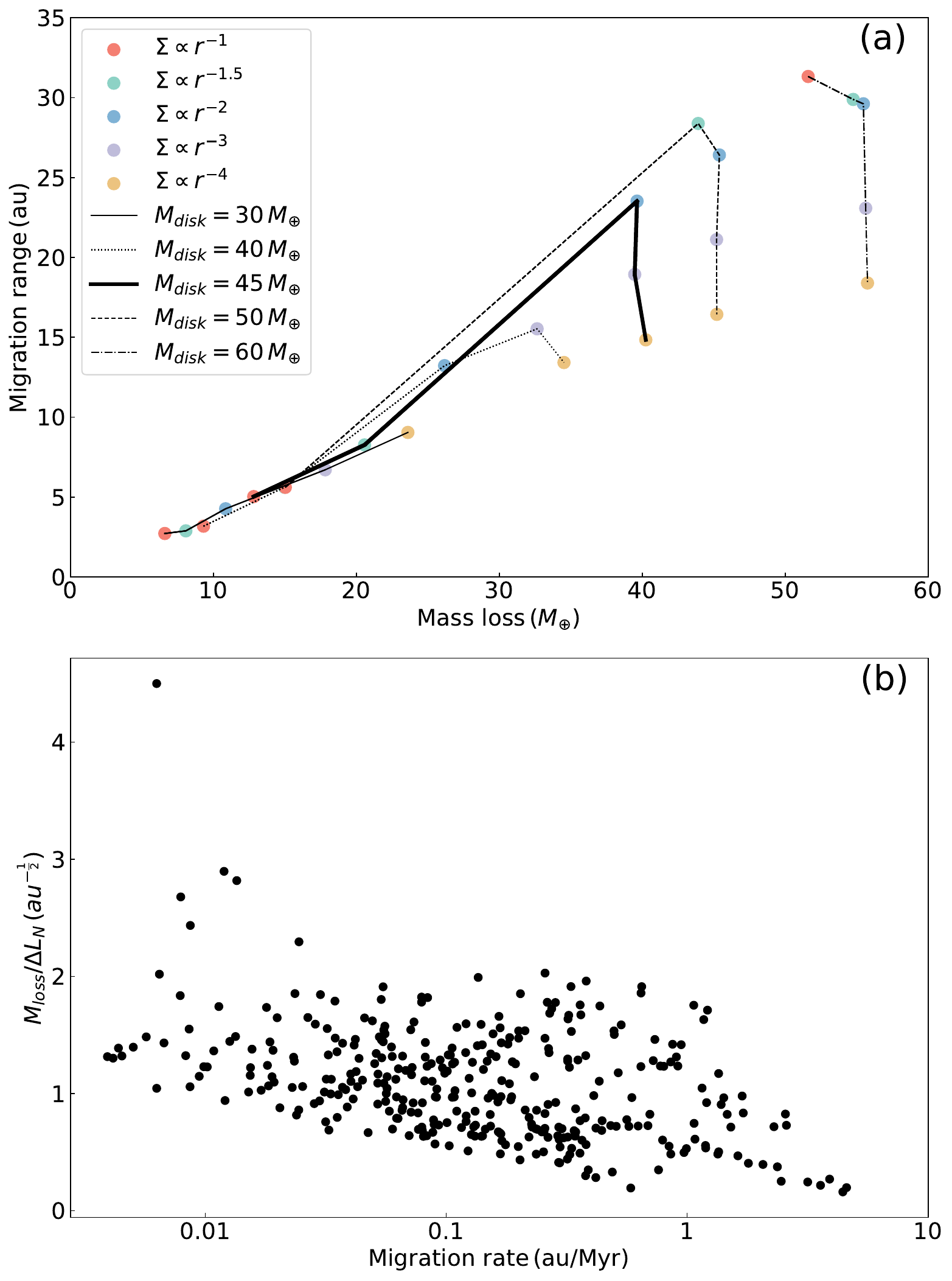}}
\caption{Neptune's migration versus the mass loss of the planetesimal disk. The upper panel illustrates the connection between the total mass loss of the planetesimal disk (horizontal axis) and the total migration distance of Neptune (vertical axis). Different colors represent different $\alpha$, while different line styles indicate different $M_{disk}$. In the lower panel, we calculated the time interval takes for Neptune to migrate per 1\,au (from 18\,au). The horizontal axis displays the average migration rate during each interval (logarithmic scale), and the vertical axis represents the ratio of consumed planetesimal mass to the angular momentum gained by Neptune within this interval.}
\label{fig:Massloss}
\end{figure}

An issue of greater concern arises from how different planetesimal densities may affect Neptune's migration. Previous studies have provided some empirical conclusions, generally suggesting that migration rate is proportional to the mass of the planetesimal disk \citepads[e.g.,][]{Kirsh2009}. However, this relationship may break down when Neptune sweeps through regions of different densities or migrates at different rates. In this study, we focus on the consumption of small bodies and converted it into the mass loss of the planetesimal disk. The vast majority of small bodies are removed because they are scattered out of the Solar System by Jupiter, but before that most of them have to be transported inward to Jupiter's vicinity via Neptune's scattering. Therefore, the loss of planetesimals reflects the mass of planetesimals encountered by Neptune. Fig.~\ref{fig:Massloss} illustrates the relationship between planetesimal loss and Neptune's migration rate.

In the Fig.~\ref{fig:Massloss}a, the relationship between the total mass loss of the planetesimal disk and the total migration distance of Neptune is shown. We define the ratio between the migration range and mass loss as the ``migration efficiency,'' which measures the distance Neptune migrates per Earth mass of planetesimals consumed.

In the bottom left corner of the Fig.~\ref{fig:Massloss}a, nine points correspond to simulations where the $R_{2N:1U}$ was not crossed. These nine simulations have almost the same migration efficiency of $0.38\pm0.02\,au/M_{\oplus}$, meaning that each Earth mass of planetesimals leads to Neptune migrating by 0.38 au. As $M_{disk}$ or $\alpha$ increases, we observe that Neptune migrates further, while the migration efficiency also noticeably improves. Among these, the simulation with $M_{disk} = 50\,M_{\oplus}$ and $\Sigma \propto r^{-1.5}$ stands out, with a migration efficiency of 0.65 $au/M_{\oplus}$, which indicates a significant reduction in the required planetesimal mass for the same migration distance.

As was mentioned earlier, if $\alpha$ increases further, the total migration distance will decrease due to the lack of material in the outer disk to sustain migration. At the same time, the mass loss will slightly increase, leading to a decrease in migration efficiency. Among them, the simulation with $M_{disk} = 60\,M_{\oplus}$ and $\Sigma \propto r^{-4}$ has the lowest migration efficiency: only 0.33\,$au/M_{\oplus}$. We speculate that in a disk with excessively high density, the randomness of close encounters increases substantially. In other words, the probability of Neptune losing angular momentum during close encounters with planetesimals is higher in an over-dense disk, which leads to the unnecessary consumption of planetesimals. Thus far, we conclude that migration efficiency is relatively low when the disk density is either too small or too large. In contrast, a planetesimal disk with a moderate density yields the highest migration efficiency and is the most ``economical'' in terms of planetesimal consumption.

Since Neptune may go through different stages within a single simulation, we can also observe the relationship between Neptune's migration rate and the planetesimal mass loss at different stages. Therefore, starting from 18\,au, we measured the time required for each 1\,au increment of Neptune's migration. We calculated Neptune's migration rate ($\dot{a}_{N}$), its angular momentum change ($\Delta L_N$), and the mass loss of planetesimals ($M_{loss}$), and plot their relationship in the Fig.~\ref{fig:Massloss}b. In the calculation of $\Delta L_N$, both the gravitational constant and solar mass are set to unity; therefore, the angular momentum of Neptune is given by $M_N \sqrt{a_N(1-e_N^2)}$.

The outlier visible in the upper left corner of Fig.~\ref{fig:Massloss}b originates from the $\alpha=4$ and $M_{disk}=60\,M_{\oplus}$ simulation. In this case, Neptune initially migrated too rapidly, then while moving from 43\,au to 44\,au it exhibited temporary inward motion before resuming outward migration. This unusual behavior consumed approximately $7\,M_{\oplus}$ of planetesimals for only 1\,au of net migration. We have excluded this data point from subsequent analysis.

For the remaining points, we note that the $M_{loss}/\Delta L_N$ gradually decreases as the migration rate increases. By fitting the trend of these data points, the $M_{loss}/\Delta L_N$ is roughly proportional to ${\dot{a}_N}^{-0.155}$. Specifically, although the absolute value of the exponent ($-0.155$) is small, it implies that a tenfold increase in migration rate reduces the required planetesimal mass consumption by approximately 30\% for the same angular momentum change of Neptune. We emphasize that this proportional relationship only provides a rough trend description and cannot reliably predict specific planetesimal consumption.

In the lower right corner of the Fig.~\ref{fig:Massloss}b, several data points exhibit exceptionally low $M_{loss}/\Delta L_N$ values. These cases, with migration rates of approximately 4\,au/Myr, achieve remarkably high migration efficiencies of $\sim 2\,au/M_{\oplus}$, which are much higher than the values derived from the Fig.~\ref{fig:Massloss}a. One plausible explanation is that during extremely rapid migration, a significant fraction of small bodies may infiltrate Neptune's inner orbital region without undergoing scattering interactions with the planets, as was described by \citetads{Gomes2004}. Perhaps these objects could play a role in the migration of Neptune, but they would not be consumed immediately in the process. As these points approach the time-resolution limit of our simulations, further investigation is needed to verify these findings.

\subsection{Termination of planetary migration}

Since the planetary migration is attained by the consumption of planetesimals, we can hypothesize that Neptune's migration will continue until the planetesimals become too sparse to maintain it. Given the final stopping position of Neptune, we can calculate the initial surface density at that location with Eq.~\ref{eq:surface_density}. To ensure that Neptune's final position is not artificially influenced by the disk boundaries, we excluded five simulations where the migration distances were too short ($a_f<21\,au$) or too long ($a_f>45\,au$). The results of the remaining simulations are summarized in Table \ref{tab:an_prediction}.

\begin{table}[htb!]
        \caption{Initial surface density of planetesimals at the location where Neptune stops in different simulations.}
	\centering
	\begin{tabular}{cccc|c}
	 $M_{disk}$ & $\alpha$ & $a_{f}$ & $\Sigma(a_{f})$ & Migration condition  \\ 
	 $(M_{\oplus})$ &      & $(au)$  & $(M_{\oplus}/au^2)$ &   \\ 
     \hline
     
		30 & 1.5 & 21.304        & 0.00858        &  \multirow{10}{*}{}\\
		30 & 2   & 22.408        & 0.00929        &                    \\
		30 & 3   & 25.235        & 0.00836        &  \\
		40 & 1   & 21.672        & 0.00918        &  $R_{2N:1U}$ not crossed  \\
		40 & 1.5 & 24.117        & 0.00950        &  0.00924 $\pm$ 0.00058                 \\
		45 & 1   & 23.274        & 0.00962        &                    \\
		45 & 1.5 & 26.920        & 0.00906        &                    \\
		50 & 1   & 24.048        & 0.01034        &                    \\ \hline
		30 & 4   & 27.828        & 0.00593        &  critical state    \\
	  40 & 2   & 31.674        & 0.00621	    &                    \\ \hline
		40 & 3   & 33.950        & 0.00458        & \multirow{10}{*}{} \\
		40 & 4   & 32.188        & 0.00442        &                    \\
		45 & 2   & 41.998        & 0.00397        &                    \\
		45 & 3   & 37.493        & 0.00382        &                    \\
		45 & 4   & 34.319        & 0.00384        &   $R_{2N:1U}$ crossed  \\
		50 & 2   & 44.817        & 0.00388        &   0.00385 $\pm$ 0.00037\\
		50 & 3   & 39.941        & 0.00351        &                    \\
		50 & 4   & 36.019        & 0.00352        &                    \\
		60 & 3   & 42.083        & 0.00360        &                    \\
		60 & 4   & 38.149        & 0.00336        &                    \\ \hline

	\end{tabular}
	 \tablefoot{From left to right, the four columns represent the $M_{disk}$, the $\alpha$, the final position of Neptune, $a_{f}$, and the initial surface density, $\Sigma(a_{f})$, of the disk at that location. In the rightmost column, all simulations are categorized into three groups, and the average values and standard deviations of $\Sigma(a_{f})$ are given for each group.}
	\label{tab:an_prediction}
\end{table}

In Table \ref{tab:an_prediction}, based on whether the $R_{2N:1U}$ is crossed, we divided the remaining 20 simulations into three groups: 8 for not crossed, 10 for crossed, and 2 on the critical state. Within each group, the initial surface density at Neptune's final position ($\Sigma(a_{f})$) is relatively constant. When the $R_{2N:1U}$ is not crossed, the average density, $\Sigma(a_f)$, is 0.00924\,$M_{\oplus}/au^2$, while it is 0.00385\,$M_{\oplus}/au^2$ if the resonance is crossed, 2.4 times smaller than in the former case. This indicates that while the crossing of $R_{2N:1U}$ may not lead to significant changes in the $a_U$ or $a_N$ directly, once this crossing occurs, it can prolong Neptune's migration significantly, continuing until a much lower disk density location.

As for the current state of the real Solar System, the period ratio between Uranus and Neptune is about 1.96, very close to but not exactly in $R_{2N:1U}$. This, on the one hand, implies that the surface density of the planetesimal disk near 30\,au might have been $\sim 0.009\,M_{\oplus}/au^2$ in the early Solar System. On the other hand, it also suggests that Neptune's current position is a very delicate balance: if there is a slight increase in the planetesimal material, Neptune would cross $R_{2N:1U}$ and migrate to a much more distant location.

\subsection{Location of $R_{2N:1U}$ crossing}

According to Table \ref{tab:an_prediction}, the migration distance of Neptune would be significantly extended after crossing $R_{2N:1U}$, which clearly does not align with the actual configuration of the Solar System, in which Neptune has not crossed $R_{2N:1U}$. In the previous simulations, Neptune crosses the $R_{2N:1U}$ at approximately 26-27\,au. This occurs because Uranus tends to occupy a more inward position in our model compared to its actual location in the Solar System. This issue was previously identified by \citetads{Gomes2004}, and has been resolved within the framework of giant planet instability models. For instance, in the original Nice model \citepads{Tsiganis2005}, the order of Neptune and Uranus was swapped during evolution. \citetads{Nesvorny2012} demonstrated that this problem could also be explained by introducing a fifth or sixth giant planet.

Different versions of the Nice model vary in their specifics, but the evolution after the giant planet instability should be similar; that is, the planets should undergo a steady migration driven by the planetesimals until reaching their current position. Although the simulations in this paper do not involve giant planet instability, the migration paths of giant planets after the instability can be traced back by these simulations, thereby helping to constrain the occurrence of giant planet instability. 

To this end, we conducted a hypothetical experiment to simulate the continuation of migration in the real Solar System. We used the current semimajor axes of the major planets [5.2\,au, 9.6\,au, 19.3\,au, and 30.3\,au] as $A_{ini}$. A total of 9000 small bodies, each with a mass of 0.005\,$M_{\oplus}$, were distributed between 30.5\,au and 61\,au following a surface density profile of $\Sigma \propto r^{-1.5}$. The simulation confirmed that the current Solar System is in a highly delicate state. With only a slight additional migration, Neptune would cross the $R_{2N:1U}$ when $a_N$ and $a_U$ is 30.7\,au and 19.3\,au, respectively. It is noteworthy that \citetads{Gomes2016} demonstrated the possibility that Neptune was once closer to $R_{2N:1U}$ than it is now, which could explain the scarcity of Neptune Trojans. This finding implies that Neptune may have never actually reached or crossed the $R_{2N:1U}$, as such an event would have completely eliminated its Trojan population.

Below, we explore the extent to which the architecture of the real Solar System can be replicated without introducing giant planet instability. It is natural to consider that placing Uranus initially at a more distant position is an effective strategy. On the other hand, according to the principle that the planets tend to spread apart during migration \citepads[][]{Hahn1999, Gomes2004}, if Uranus is positioned too close to Neptune, its outward migration would instead be suppressed. For a similar reason, placing Neptune further outward is another viable method, as this would reserve some ``space'' for Uranus to migrate outward. It should be noted that Neptune's radial migration must have exceeded $\sim$5\,au to reproduce Pluto's current orbital eccentricity \citepads{Malhotra1995}, and excite the inclinations of TNOs to observed levels \citepads{Nesvorny2015b}.

\begin{figure}[!htb]
\centering
\resizebox{\hsize}{!}{\includegraphics{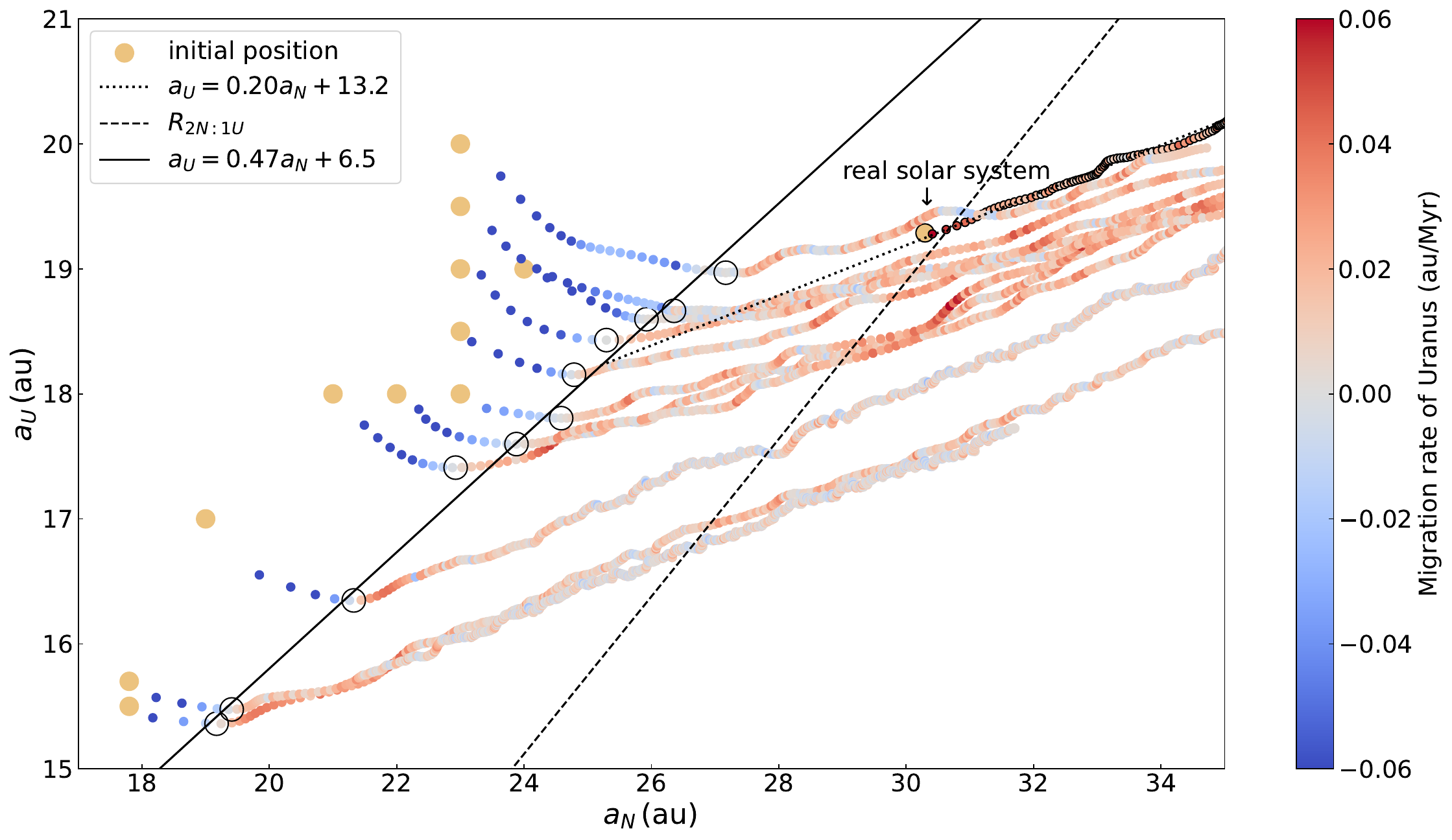}}
\caption{Evolution of the positions of Uranus and Neptune. The horizontal and vertical axes represent the $a_N$ and $a_U$, respectively. The color of the points indicates the migration rate of Uranus at each position. The orange dots denote the initial $a_N$ and $a_U$ in each set of simulations. The points with black borders represent hypothetical extensions of the real Solar System's migration, while the dotted black line shows the linear fit of these points with black borders. The hollow circles mark the positions where $a_U$ reaches its minimum value, and the solid black line represents the linear fit of all black circles. The dashed black line indicates the location corresponding to the $R_{2N:1U}$.}
\label{fig:Res_loc}
\end{figure}

Therefore, we fixed the $a_J$ and $a_S$ at 5.45\,au and 8.8\,au, respectively, and conducted several simulations by adjusting the initial $a_N$ and $a_U$. In these simulations, 9000 small bodies were distributed from $a_N$ + 0.2\,au to 50\,au, each with a mass of 0.005\,$M_{\oplus}$ and the surface density profile following $\Sigma \propto r^{-1.5}$. After appropriate smoothing, we present the evolutionary trajectories of Uranus and Neptune in Fig.~\ref{fig:Res_loc}. We tested multiple combinations of initial $a_N$ and $a_U$, corresponding to the orange dots in Fig.~\ref{fig:Res_loc}.

In Fig.~\ref{fig:Res_loc}, it can be observed that due to the adjustment of Uranus's initial position further outward, Uranus initially undergoes inward migration in all simulations. This is the most significant difference compared to the simulations in Section 3.1 (see Fig.~\ref{fig:PNU}). After migrating inward for a certain distance, Uranus' migration reaches a turning point, which have been marked with hollow circles in Fig.~\ref{fig:Res_loc}. An interesting phenomenon is that all these circles can be represented well by a linear relationship, namely $a_U = 0.45\,a_N + 6.7\,au$. In Fig.~\ref{fig:Res_loc}, we use a solid black line to represent this fit linear relationship. This line roughly divides the migration evolution into two phases: in the upper left region of the solid black line, Uranus migrates inward, while in the lower right region, its migration is outward.

Once the migration evolution crosses the turning point and enters the lower right region of the solid black line, it can be observed that the outward migration of Uranus and Neptune is largely proportional. In other words, the two giant planets evolve along a straight line toward the upper right. The slope of these lines varies from 0.06 to 0.2, with steeper slopes generally occurring when Uranus and Neptune are farther apart (i.e., at lower positions in Fig.~\ref{fig:Res_loc}).

Fig.~\ref{fig:Res_loc} can help evaluate whether the simulated Neptune and Uranus can evolve into the locations resembling those observed in the actual Solar System. It is evident that the simulation representing the hypothetical extended migration of the real Solar System is located in the upper part of Fig.~\ref{fig:Res_loc}, indicating that the real Neptune and Uranus are more compact compared to most of the simulations. This proximity is crucial for delaying the occurrence of the $R_{2N:1U}$ until $a_N$ is beyond 30\,au.

On the other hand, we can also use this proportional relationship to infer the migration history in the real Solar System. Through a simple fitting, it can be roughly determined that Neptune and Uranus evolved along the line $a_U = 0.20\,a_N + 13.2\,au$ (the dotted line in Fig.~\ref{fig:Res_loc}). In the early Solar System, the two planets could have started from a certain point near this line and evolved to their current positions. However, this line intersects with the line corresponding to the turning point of Uranus's migration direction (the solid black line in Fig.~\ref{fig:Res_loc}) at $a_N = 24.8\,au$ and $a_U = 18.2\,au$. 

Notably, the slope of the dotted line is steeper than those of neighboring simulations. This particular behavior likely stems from specific disk parameters ($M_{disk} = 45\,M_{\oplus}$ and $\alpha = 1.5$) and may change with the initial conditions. As is discussed in Section 3.1, in relatively massive disks, the $\Delta a_U/\Delta a_N$ tends to be greater. On the other hand, the slope is relatively low when $a_U$ reaches its minimum value. Under the influence of these factors, the actual intersection point is likely further outward, probably near $a_N \approx 26\,au$ and $a_U \approx 18.7\,au$.

Dynamical constraints require that the initial $a_N$ was interior to 25\,au \citepads{Malhotra1995,Nesvorny2015b}, which means any reconstruction of the ice giants' orbital history that excludes giant planet instability will inevitably include an inward migration phase for Uranus. Based on our ensemble of simulations, this early migration phase exhibits considerable uncertainty. The minimum viable $a_N$ is approximately 23 to 24\,au and the $a_U$ is about 20\,au at this point. Further backtracking becomes dynamically implausible as it would place the two ice planets in an unrealistically close configuration.

The preceding analysis assumes no giant planet instability has occurred. Crucially, Neptune's minimum migration distance of $\sim$5\,au implies that its initial configuration with Uranus must have been exceptionally compact, requiring Uranus to execute initial inward migration followed by outward migration. This constrains their primordial orbits to a narrow space that $a_N \approx 24\,au$ and $a_U \approx 20\,au$. However, this configuration still presents a dynamical anomaly, with the Neptune-Uranus separation <5\,au, while there is a wide gap of more than 10\,au between Saturn and Uranus. This inconsistency strongly suggests that giant planet instability must have occurred.

Under the framework of the Nice model, recent studies indicate that the giant planet instability occurred relatively early \citepads[e.g.,][]{Clement2018, Nesvorny2018b, Clement2019, Ribeiro2020}. Following the instability, Neptune and Uranus could have been in a relatively small separation. Our results can also help to provide constraints on the post-instability orbital configurations of Uranus and Neptune. As was established earlier, the post-instability state of ($a_U$, $a_N$) is likely to initiate near a certain point on the arc-shaped trajectory extending from ($\sim$20\,au, $\sim$24\,au) through ($\sim$18.7\,au,$\sim$26\,au) to (19.3\,au, 30.3\,au), and subsequently evolving along it. The inferred trajectories of $a_N$ and $a_U$ are sensitive to the assumed initial conditions, and thus carry substantial uncertainty. Future studies will be needed to provide better constraints on this evolutionary pathway. However, regardless of initial conditions, during the post-instability phase, the planetesimal-driven migration distance of Neptune is unlikely to be larger than $\sim$7\,au.

\section{Migration in a planetesimal disk with ring}

Previous studies have proposed that planetesimals form through the streaming instability \citepads[e.g.,][]{Youdin2005, Cuzzi2008, Simon2016}, a process potentially triggered near snow lines or sublimation lines of particular materials \citepads[e.g.,][]{Izidoro2022, Morbidelli2022}. This mechanism naturally produces planetesimal disks with ring-and-gap structures, where the dense rings play a crucial role in the formation and growth of planets \citepads{Lichtenberg2021, Izidoro2022, Morbidelli2022, Batygin2023}. Due to the distinct snow line locations of different volatile materials, these ring structures are likely to be abundant and may even exist beyond Neptune. Possibly, these outermost rings are sparse that persisted until the planetary migration epoch. While some of these outer planetesimals may have grown to Pluto-sized objects, they have not accreted into planets.

Several studies have supported the existence of numerous Pluto-sized bodies in the trans-Neptunian region. For example, \citetads{Nesvorny2016} put forward that thousands of such objects could have caused Neptune's migration to be grainy, and thereby reduced the capture efficiency of resonant TNOs. On the other hand, the biggest satellite of Neptune, Triton, evolving in a retrograde orbit, is assumed to be a captured object. However, because the capture efficiency is very low \citepads[e.g.,][]{Stern1991,Nesvorny2007,Vokrouhlicky2008,Nesvorny2014}, its existence implies that there may be a large number of objects as large as Triton.

An alternative scenario is that planetesimal rings could have formed during the giant planet instability. When Neptune experiences substantial orbital jumps, objects previously trapped in the MMRs may be partially liberated. These released objects would cease co-migrating with Neptune, instead forming a ring-like structure. This mechanism has been demonstrated by \citetads{Nesvorny2015a} to explain the origin of the cold classical kernel.

In the previous section, we have already known that crossing the MMR between Uranus and Neptune can accelerate migration. Additionally, since migration itself requires a continuous supply of planetesimals, it is reasonable to hypothesize that encountering a relatively dense planetesimal ring may also help accelerate the planetary migration. However, we emphasize that the existence of such rings requires additional evidence, and more detailed dynamical studies are needed to fully characterize their properties. In this work, we only adopt a simplified ring model, and try to understand its fundamental effects on planetary migration dynamics.

\subsection{Planetesimal ring far from MMR}

Initially, we aimed to isolate the influence of MMR and focus solely on the effect of a planetesimal ring. In the previous section, we have shown that if the $A_{ini}$ are set as [5.45\,au, 8.7\,au, 15.5\,au, and 17.8\,au], Neptune will cross $R_{2N:1U}$ around 26 to 27\,au and trigger subsequent acceleration. To avoid interference from $R_{2N:1U}$, we decided to let Neptune encounter the planetesimal ring after crossing $R_{2N:1U}$. Based on the simulation from Section 3.1 with $M_{disk}=50\,M_{\oplus}$ and $\alpha=2$, we added an additional planetesimal ring to study its effect on migration acceleration. The planetesimals within the ring have the same mass as those in the rest of the disk, which is 0.005 $M_{\oplus}$. By adjusting the number of planetesimals, the total mass of the ring is set to 2.5\,$M_{\oplus}$, 5\,$M_{\oplus}$, and 10\,$M_{\oplus}$, and these additional small bodies follow a Gaussian distribution centered at 34.5\,au with a standard deviation of 0.5\,au.

\begin{figure}[!htb]
\centering
\resizebox{\hsize}{!}{\includegraphics{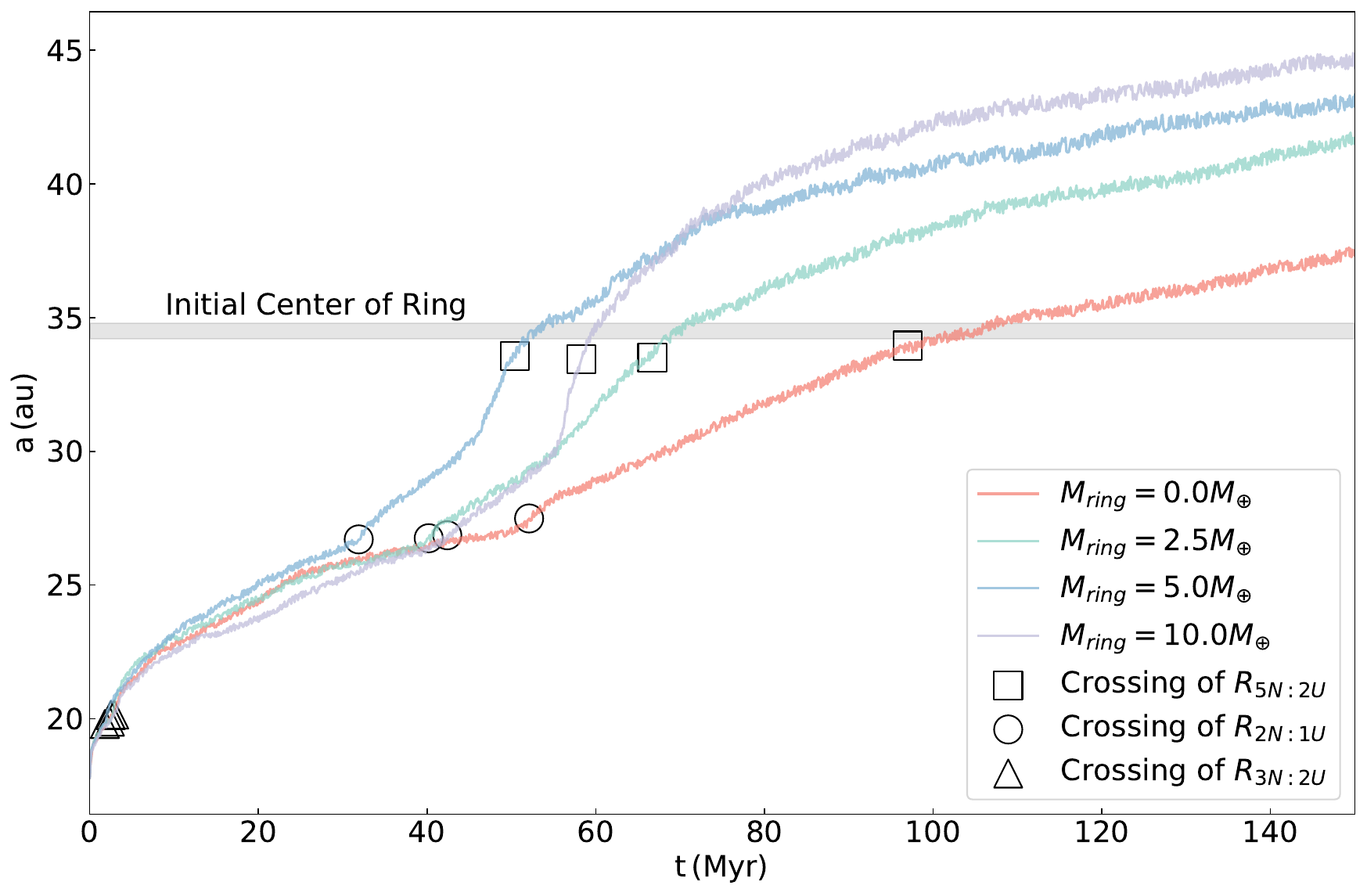}}
\caption{Evolution of Neptune with the presence of a planetesimal ring. The horizontal and vertical axes represent time and $a_N$, respectively. Lines of different colors correspond to different planetesimal ring masses. The horizontal gray line represents the initial center of the planetesimal ring (34.5\,au). The hollow triangles, circles, and squares mark the positions where Neptune crosses $R_{3N:2U}$, $R_{2N:1U}$, and $R_{5N:2U}$, respectively.}
\label{fig:Acc_ring}
\end{figure}

In Fig.~\ref{fig:Acc_ring}, we depict the migration of Neptune, where the red line is directly taken from the simulation in Fig.~\ref{fig:Nep_mig} with $M_{disk}=50\,M_{\oplus}$ and $\alpha=2$. The most notable feature in Fig.~\ref{fig:Acc_ring} is that after adding the planetesimal ring, Neptune experiences acceleration during its migration. This demonstrates that the presence of the ring can trigger Neptune's acceleration, which is independent of the previously mentioned acceleration caused by the MMR crossing.

In addition, although the initial planetesimal ring is located at 34.5\,au (indicated by the thick gray line in Fig.~\ref{fig:Acc_ring}), it is evident that Neptune's acceleration begins around 30\,au, with the maximum rate reached near\,32 au. Previous studies have shown how Neptune interacts with planetesimal disks, even in absence of direct encounters \citepads[e.g.,][]{Levison2011, Deienno2017}.  If we view the ecliptic plane from the north ecliptic pole, Neptune effectively clears out the region within a few astronomical units beyond its orbit. Moreover, the planetesimal ring does not maintain its structure but slightly disperses before Neptune arrives. As a result, Neptune will begin interacting with the planetesimal ring and accelerating as it approaches around 30\,au.

\citetads{Deienno2017} investigated the minimum separation required between Neptune and the inner edge of the planetesimal disk to maintain planetary system stability. Similarly, the location where Neptune begins to respond to the gravity of the ring may also be a complex problem. This process possibly depends on multiple factors, such as the surface density of the disk and the resolution of the simulation, as has been demonstrated by \citetads{Deienno2017}. As is shown in Fig.~\ref{fig:Acc_ring}, more massive rings tend to induce acceleration at slightly lower heliocentric distances. However, this trend is not significant, and may be influenced by our assumption of a Gaussian distribution for the ring.

The initial position of the planetesimal ring at 34.5\,au was also carefully chosen to avoid potential MMRs, $R_{2N:1U}$ and $R_{5N:2U}$. Based on Fig.~\ref{fig:Acc_ring} and Fig.~\ref{fig:PNU}, we can reasonably conclude that $R_{5N:2U}$ is unlikely to trigger acceleration in the same manner as $R_{3N:2U}$ and $R_{2N:1U}$.

When the planetesimal ring has masses of 10, 5, and 2.5\,$M_{\oplus}$, the maximum migration rates near 32\,au are approximately 1.21, 0.80, and 0.43\,au/Myr, respectively. For comparison, in simulations without a planetesimal ring, Neptune's migration rate at this location is about 0.15\,au/Myr. As a comparison, in the four simulations, Neptune's migration rate increased from approximately 0.05\,au/Myr to 0.3\,au/Myr following its crossing of $R_{2N:1U}$.

\subsection{Planetesimal ring near MMR}

\begin{figure}[!htb]
\centering
\resizebox{\hsize}{!}{\includegraphics{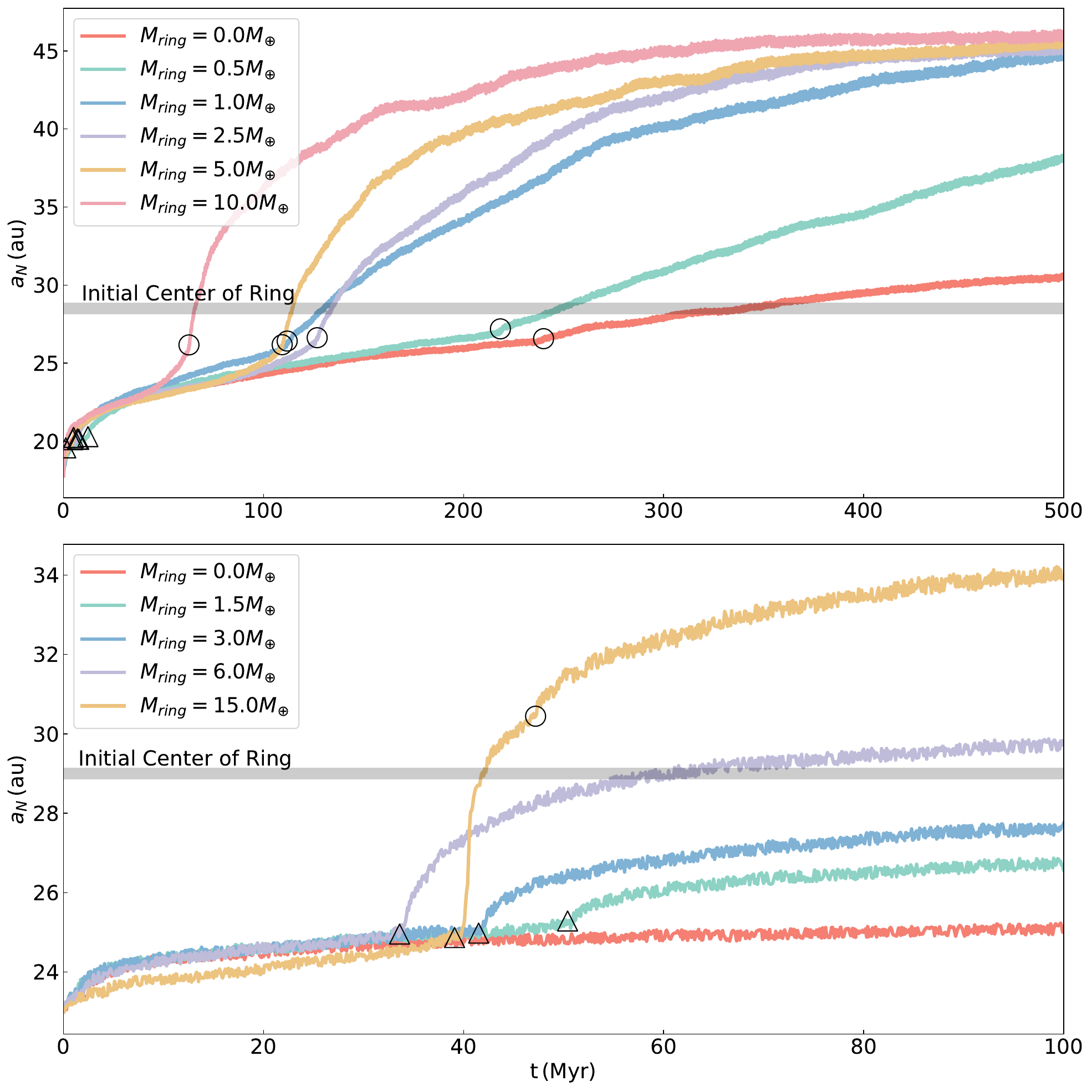}}
\caption{Same as Fig.~\ref{fig:Acc_ring} but the migration acceleration of Neptune is due to the combined effects of the planetesimal ring and MMR crossings. The initial conditions for the upper and lower panels differ as specified in Table.~\ref{tab:acc_ini}. Neptune passes through the $R_{2N:1U}$ (upper panel) or $R_{3N:2U}$ (lower panel) when it is accelerated by the planetesimal ring.}
\label{fig:Nep_acc}
\end{figure}

In the previous subsection, we discussed the acceleration effect on planetary migration induced by the planetesimal ring independently. Here we further explore the phenomena when such effects coincide with the acceleration triggered by the crossing of MMRs. For this purpose, we conducted two major sets of tests: in the first set, the initial conditions for the giant planets were adopted from Section 3.1. Building upon Section 3.4, we modified the initial conditions in our second set to better match the current semimajor axes of the giant planets (notably, other key indicators such as planetary eccentricities remain inconsistent with observed values). In these two sets of simulations, Uranus and Neptune cross the $R_{2N:1U}$ and $R_{3N:2U}$, respectively. Each set of tests had its own specific configurations, as is outlined in Table.~\ref{tab:acc_ini}, and their results are presented in Fig.~\ref{fig:Nep_acc}.

\begin{table}[htb!]
        \caption{Initial conditions used in the two sets of simulations investigating the combined acceleration effects of the planetesimal ring and MMR crossing.}
	\centering
	\begin{tabular}{cc|c|c}
	   & & Set one & Set two  \\ \hline
	\multicolumn{2}{c|}{$A_{ini}\,(au)$} & [5.45, 8.7, 15.5, 17.8]   & [5.45, 8.8, 19.8, 23] \\
    \multicolumn{2}{c|}{particle mass} & 0.005\,$M_{\oplus}$ & 0.003\,$M_{\oplus}$ \\ 
    \multicolumn{2}{c|}{MMR crossed}        & $R_{2N:1U}$ & $R_{3N:2U}$ \\ 
    \multicolumn{2}{c|}{Time}          & 500\,Myr & 100\,Myr \\ 
    \hline
        \multirow{3}{*}{} & Range    & [18\,au, 50\,au] & [23.2\,au, 50\,au] \\
       Disk               & $\alpha$ & 1.5              & 1.5 \\
                          & Amount   & 9000             & 5000 \\
        \hline
        \multirow{4}{*}{} & Center & 28.5\,au & 29\,au \\
       Ring               & $\sigma$ & 0.5\,au & 0.5\,au \\
                          & \multirow{2}{*}{Amount}   & 0, 100, 200, 500 & 0, 500, 1000 \\
                          &                           & 1000, 2000       & 2000, 5000 \\
    \hline

	\end{tabular}
	\tablefoot{The planetesimals in the ring are generated according to a Gaussian distribution. Each planetesimal has the same mass, and the numbers in the disk and the ring are listed. The initial masses of the planets, as well as the initial eccentricities and inclinations of the planets and planetesimals, remain the same as the ones described in Section 2 and are not listed again.}
	\label{tab:acc_ini}
\end{table}

As is shown in the upper panel of Fig.~\ref{fig:Nep_acc}, with the addition of the planetesimal ring, Neptune's acceleration can occur earlier and faster. As was mentioned above, Neptune has a gravitational influence range of about 4\,au, so in some simulations, Neptune experiences a slight acceleration before crossing $R_{2N:1U}$ , especially for the heavier rings. Subsequently, the migration rate further increases after crossing $R_{2N:1U}$. Due to the higher density in the planetesimal ring, Neptune encounters additional material during its acceleration, allowing it to migrate rapidly to more distant locations.

For relatively massive planetesimal rings ($M_{ring} \geq 1\,M_{\oplus}$), Neptune's migration will ultimately extend to around 46\,au. According to Eq.~\ref{eq:surface_density}, the initial planetesimal ring density at 46\,au is approximately 0.0036\,$M_{\oplus}/{au}^2$, which aligns with the conclusions in Table \ref{tab:an_prediction} when migration crosses $R_{2N:1U}$. This also implies that in this scenario, the final position of Neptune is still largely determined by the density distribution profile of the planetesimal disk, rather than the total amount of planetesimals encountered during Neptune's migration history.

The positive feedback mechanism makes the timing of resonance crossing equally critical in shaping planetary migration outcomes. For example, although Neptune can cross $R_{2N:1U}$ after 200\,Myr even without the addition of a planetesimal ring, by this time, the planetesimals near Neptune have been significantly depleted, limiting further migration. This indicates that even minor differences in the planetesimal disk ($M_{ring}\leq 1\,M_{\oplus}$) can lead to an advance in the timing of Neptune crossing $R_{2N:1U}$, thereby greatly affecting Neptune's final position.

In the lower panel of Fig.~\ref{fig:Nep_acc}, the migration acceleration is sharply and abruptly initiated, and is almost entirely marked by crossing $R_{3N:2U}$ as the starting point. In cases with a lower planetesimal ring mass ($M_{ring} \leq 3\,M_{\oplus}$), Neptune does not even reach 29\,au by the end of the simulation. 

In contrast to the first set of simulations, in the second set, due to the overall lower density of the planetesimal disk and the higher mass contribution of the ring, Neptune's final position is primarily determined by the mass of the planetesimal ring. In simulations without the planetesimal ring, Neptune only migrated to approximately 25\,au, where the initial surface density was $\sim0.0042\,M_{\oplus}/{au}^2$. This outcome is not comparable with cases in Table \ref{tab:an_prediction} in which $R_{2N:1U}$ was not crossed, but does not contradict its conclusions, because the surface density here is never sufficiently high to drive the migration, even in the innermost regions. For $M_{ring} = 15\,M_{\oplus}$, Neptune migrated outward to about 34\,au, where the initial surface density was only $\sim0.0027\,M_{\oplus}/{au}^2$, which is lower than all cases presented in Table \ref{tab:an_prediction}. This suggests that in a planetesimal disk with a ring structure, the mass of the ring, as well as the gaps between rings \citepads[e.g.,][]{ALMA2015}, could play a crucial role in influencing Neptune's final position.

Through the simulations in this section, it has been recognized that the planetesimal ring cannot only independently drive migration acceleration but can also combine with the crossing of MMRs. The changes of migration rates induced by the planetesimal ring tend to be relatively smooth, while resonance crossings can lead to abrupt changes in migration rates. When both mechanisms occur simultaneously, resonance crossing acts as a trigger, with the planetesimal ring providing the material supply necessary for migration. Combined with the positive feedback mechanism inherent in migration, the mass of the planetesimal ring does not need to be large to significantly alter the migration process, affecting Neptune's final position.

\section{Conclusion and discussion}

Previous works have shown that in either a planetesimal disk or a gas disk, planetary migration exhibits positive feedback mechanisms \citepads[see e.g.,][]{Ida2000,Masset2003,Peplinski2008}, theoretically making migration acceleration possible. On the other hand, more and more research suggests the presence of ring structures in dust or gas disks\citepads[see, e.g.,][]{Izidoro2022, Morbidelli2022, Batygin2023}. Within these rings, planetesimals are more densely distributed and may form in a more efficient way. These ring structures may initially originate from the snow lines or sublimation lines of specific materials in the early Solar System \citepads[e.g.,][]{Izidoro2022}, or they may result from the clearing of surrounding material during subsequent planetary formation processes \citepads[e.g.,][]{ALMA2015}. It is reasonable to assume that during planetary formation, most of the surrounding material would have been accreted by planets or rapidly scattered to distant regions. But at relatively distant locations, planetary formation proceeds more slowly, and it is plausible to have a ring composed of planetesimals. It is thus natural to infer that planetary migration acceleration might be invoked by these ring structures.

In this paper, to investigate the acceleration mechanism of planetary migration, we conducted detailed numerical simulations. We first examined the case of planetary migration in a plain planetesimal disk (Fig.~\ref{fig:Nep_mig}). We observed that whenever Neptune crossed $R_{2N:1U}$ or $R_{3N:2U}$, the migration rate significantly increased (Fig.~\ref{fig:PNU}). Since Neptune and Uranus did not remain in the MMR, we propose that the MMRs are not the driving factors for accelerating migration but rather act as triggers.

By studying the relationship between Neptune's migration and planetesimal consumption (Fig.~\ref{fig:Massloss}), we confirmed the existence of a positive feedback mechanism: faster migration consumes planetesimals more efficiently, while slower migration gradually empties the nearby region of Neptune, making it more challenging to sustain the migration. Specifically, when Neptune's migration rate increases by an order of magnitude, the mass of planetesimals consumed per unit of angular momentum change of Neptune decreases by approximately 30\%. We found that the MMR crossing greatly extends the migration distance because of the positive feedback mechanism (Table \ref{tab:an_prediction}). 

On the other hand, this also indicates that Neptune in the real Solar System currently occupies a delicate position, as it is close to $R_{2N:1U}$. Additionally, Jupiter and Saturn are also very close to their 5:2 MMR. If the mass of the planetesimal disk in the early Solar System were slightly larger, the giant planets would have quickly crossed these resonances, causing Neptune to migrate to a much farther location. It is challenging to determine whether this is a coincidence or an inevitable outcome of evolution.

As has been proposed in previous studies \citepads[e.g.,][]{Ida2000, Gomes2004}, the runaway migration should be self-sustaining, requiring no additional planets. This process drives Neptune to the disk edge, significantly increasing its orbital separation from the other planets. Therefore, one of the key characteristics of such independent rapid migration is that it will be accompanied by an inward migration of Neptune subsequently. However, in our simulations, definitive runaway migration only occurred when the disk is the most massive ($M_{disk}=60\,M_{\oplus}$ and $\alpha=4$; see Fig.~\ref{fig:Nep_mig}). Therefore, most simulations in this paper still belong to damped migration, even when migration acceleration is triggered by MMR crossing or planetesimal rings.

The results in Table \ref{tab:an_prediction} concerning Neptune's final position can be regarded as a specific instance of the conclusion described by \citetads{Gomes2004}. In summary, if Neptune experiences damped migration, its final location depends primarily on the local initial surface density (and whether it has crossed $R_{2N:1U}$). In contrast, Neptune in a runaway migration will be propelled outward until it reaches the disk edge.

By simulating planetary migration without invoking giant planet instability, we trace back the dynamical histories of Uranus and Neptune (Fig.~\ref{fig:Res_loc}). We find that the migration of Neptune is severely constrained on this condition, which indicates the necessity of introducing the giant planet instability \citepads[e.g.,][]{Tsiganis2005, Nesvorny2012}. Furthermore, although our model employs simplified assumptions, it provides valuable constraints on the post-instability configuration of the planetary system.

In Section 4, we focused on the influence of the planetesimal ring on planetary migration. From Fig.~\ref{fig:Acc_ring}, we observed that migration acceleration driven solely by the planetesimal ring exists in the absence of MMR crossings. In the simulations, the distance of acceleration can significantly exceed the width of the planetesimal ring. Additionally, considering the dispersal of the planetesimal ring, the width of the ring may not have a significant influence compared to the total mass of the ring.

In Fig.~\ref{fig:Nep_acc}, we demonstrate the combined effects of the planetesimal ring and MMR crossings on migration acceleration. The triggering effect of MMR crossings remains pronounced, as it induces nearly instantaneous changes in migration rates, while the planetesimal ring serves to replenish material. In a relatively dense disk (upper panel of Fig.~\ref{fig:Nep_acc}), due to the positive feedback mechanism, even minor adjustments to the planetesimal ring mass can lead to significantly different simulation outcomes. In contrast, in a relatively sparse disk (lower panel of Fig.~\ref{fig:Nep_acc}), the mass of the planetesimal ring largely determines Neptune's final position.

In the lower panel of Fig.~\ref{fig:Nep_acc}, we selected $A_{ini}$ = [5.45\,au, 8.8\,au, 19.8\,au, and 23\,au] to prevent Neptune from crossing $R_{2N:1U}$ before reaching 30\,au. This design, however, leads to Neptune crossing $R_{3N:2U}$ around 25\,au. This crossing likely occurred in reality (unless $a_N$ had been beyond 25\,au after the occurrence of giant planet instability), implying that Neptune experienced a migration acceleration around 25\,au. At this point, Neptune's 1:2 MMR is located at 39$\sim$40\,au, which coincides with the current position of Neptune's 2:3 MMR -- the inner boundary of the classical population of TNOs.

As the capture efficiency of the Neptunian 1:2 resonance decreases during rapid Neptune migration \citepads{Li2023}, the existence of the cold classical kernel (a subclass of cold TNOs located near 44\,au \citepads{Petit2011}) could be explained by migration acceleration. This scenario provides an alternative to the resonant leakage mechanism \citepads{Nesvorny2015a}, which is based on a major jump of Neptune induced by giant planet instability. Both explanations share a key requirement: Neptune must undergo rapid movement near 27.8\,au, whether due to ice giant scattering or the acceleration mechanism proposed here. However, 27.8\,au does not align with the position of crossing $R_{3N:2U}$ in our simulation. As an alternative, perhaps a planetesimal ring near 32\,au could provide an explanation. In this case, Neptune only moved $\sim$2\,au from migration acceleration to eventual termination. According to the lower panel of Fig.~\ref{fig:Nep_acc}, this planetesimal ring may have a mass of around 2\,$M_{\oplus}$.

The boundary of the planetesimal disk is a major simplification in our model. We considered a disk extending to 50\,au to facilitate the study of the rate changes of planetary migration, while some previous works preferred a truncated disk \citepads[e.g.,][]{Gomes2004, Nesvorny2012, Nesvorny2015b, Deienno2017}, possibly due to photoevaporation or stellar flyby \citepads[e.g.,][]{Adams2010}. Although this results in a larger $M_{disk}$, the surface density remains comparable to prior studies. On the other hand, the inner boundary of the disk is only 0.2\,au beyond Neptune, which could enhance the early migration rate of Neptune in our simulations, though its overall effect on the final orbital configuration remains limited. Furthermore, \citetads{Deienno2017} demonstrate that Neptune's migration will be triggered instantaneously for any disk-planet separations $\lesssim3\,au$.

Additionally, in this study, the initial masses of the planets were set to the current values in the Solar System. However, during the simulations, the masses of the giant planets always experience slight growth due to the mass accretion. For example, in the simulation from Section 3.1 with $M_{disk}=50\,M_{\oplus}$ and $\alpha = 2$, the masses of the four planets grew from 317.8, 95.152, 14.536, and 17.147\,$M_{\oplus}$ to 318.2, 95.327, 14.756, and 17.832\,$M_{\oplus}$, respectively. This represents a total increase of approximately 1.5\,$M_{\oplus}$, accounting for about 3.3\% of the total mass lost from the planetesimal disk. As a result, the simulations slightly overestimate the planetary masses, which may lead to a modest reduction in the extent of planetary migration but does not affect the qualitative conclusions.

The positive feedback mechanism during migration also amplifies the instability of simulation results. By observing the crossing times of $R_{2N:1U}$ in Fig.~\ref{fig:Acc_ring} (or $R_{3N:2U}$ in the lower panel of Fig.~\ref{fig:Nep_acc}), we notice that even with identical planetesimal disk distributions (where the outer planetesimal ring has not yet taken effect), the crossing times can differ by up to about 20\,Myr. The timing of MMR crossing also affects the effectiveness of subsequent acceleration. For instance, in Section 3.1, in the simulation with $M_{disk}=40\,M_{\oplus}$ and $\alpha=2$ (or $M_{disk}=30\,M_{\oplus}$ and $\alpha=4$), although Neptune eventually crosses $R_{2N:1U}$, the time of crossing is too late for Neptune to reach its expected position (corresponding to the critical state simulation in Table \ref{tab:an_prediction}). 

In this study, we investigate how MMR crossings and planetesimal rings influence planetary migration. Both mechanisms can effectively accelerate the migration, as the migration process inherently exhibits positive feedback characteristics. However, as was previously mentioned, the positive feedback mechanism introduces fundamental unpredictability, even without artificially including giant planet instability. This implies that while we may reconstruct plausible evolutionary pathways for the Solar System, its present configuration was not an inevitable outcome. Notably, this aligns with perspectives from the Nice model framework \citepads[e.g.,][]{Nesvorny2012, Clement2021}.

\begin{acknowledgements}
        We gratefully acknowledge the anonymous referee for the valuable comments and suggestions, which have significantly improved the manuscript. This work is supported by the Science and Technology Development Fund (FDCT) of Macau (grant Nos. 0014/2022/A1, 0034/2024/AMJ, 0008/2024/AKP, 002/2024/SKL). Li-Yong Zhou thanks the support from National Natural Science Foundation of China (NSFC, Grants No.12373081 \& No.12150009) and the China Manned Space Program with grant No.CMS-CSST-2025-A16.
\end{acknowledgements}

\bibliographystyle{aa-note}
\bibliography{MigSim}
\end{document}